# CARBON NANOTUBE ARRAY VIAS FOR INTERCONNECT APPLICATIONS

Jyh-Hua Ting [1], Ching-Chieh Chiu [2], Fuang-Yuan Huang [2]

[1] National Nano Device Laboratories, No.26, Prosperity Road I, Science-Based Industrial Park, Hsinchu 300, Taiwan, R.O.C.
Phone / FAX: +886-3-5726100 ext.7720 / +886-3-5713403 E-mail: jting@mail.ndl.org.tw
[2] Department of Mechanical Engineering, National Central University, Jung-li City, Taoyuan 320, Taiwan, R.O.C.

## ABSTRACT

The material and electrical properties of the CNT single vias and array vias grown by microwave plasma-enhanced chemical vapor deposition were investigated. The diameters of multiwall carbon nanotubes (MWNTs) grown on the bottom electrode of Ta decrease with increasing pretreatment power and substrate temperature while the effects of the growth power and methane flow ratio are insignificant The decrease of CNT diameters leads to the decrease of the CNT via diode devices. The increase of growth power enhances the CNT graphitization degree and thue the conductivity of CNT via diode devices. In the same via region, the MWNT diode resistances of the array vias are lower than those of the single vias. The MWNT diode resistances on the bottom electrode of titanium are lower than those on the bottom electrode of tantalum. It may be attributed to the smaller tube diameters on the bottom electrode of Ti and the work function difference between Ta and Ti films with respect to the work function of CNTs.

## 1. INTRODUCTION

The interconnect in an integrated circuit (IC) distributes clock and other signals as well as provides power or ground to various circuits on a chip. The International Technology Roadmap for Semiconductors (ITRS) emphasizes the high speed transmission needs of the chip as the driver for future interconnect development. In general, the challenges in interconnect technology arise from both material requirements and difficulties in processing. The susceptibility of common interconnect metals to electromigration at high current densities (> $10^6$ A/cm$^2$) is a problem. On the processing side, current technology relies on three steps: dry etching to create the trenches/vias, deposition to fill metal plugs, and planarization. Aspect ratio dependent etching as well as plasma damage, cleaning, and void-free filling of high aspect ratio features are all difficult tasks.

Innovative material and processing solutions are critical to sustain the growth curve according to ITRS. Due to their high current carrying capability, high thermal conductivity, and reliability, multiwall carbon nanotubes (MWNTs) have recently been proposed as a possible replacement for metal interconnects [1, 2]. In 2013, the ITRS predicts a current density of 3.3 X $10^6$ A/cm$^2$, a value which, to date, can only be supported by carbon nanotubes (CNTs), where current densities of $10^9$ A/cm$^2$ in nanotubes without heat sinks have been reported. Although CNT array was compared to copper interconnect by modeling, the electrical characterization of the CNT array vias has not been implemented [3]. The purpose of this study is to investigate the electrical properties of the CNT single vias and array vias grown by microwave plasma-enhanced chemical vapor deposition (MPCVD).

## 2. EXPERIMENTAL METHOD

Four sets of experiments for the pretreatment power ($P_P$), growth power ($P_G$), the substrate temperature (T), and methane flow ratio [$CH_4$/($H_2$+ $CH_4$)], were conducted for CNT growth, as shown in table I. With a total flow rate of 100 SCCM [$H_2$ or $H_2$+ $CH_4$], the total pressure in the chamber was kept constant at 3999 Pa. The pretreatment and deposition times were set to be 5 and 15 minutes, respectively. Figure 1 depicts the cross section of the CNT diode structure. Two bottom electrodes of titanium (Ti) and tantalum (Ta) were adopted to examine their effects on the CNT diode resistances. For comparison, single via and array vias were designed in the same via region. 3.85 μm and 2.45 μm single vias correspond to 0.35 μm 6x6 and 4x4 array vias. Figure 2 shows the SEM images of the via patterns.

Renishaw 2000 system was utilized to obtain the Raman spectra of the CNT films. Based on the Nd-YAG laser source of 500 mW power, 532 nm-wavelength laser pulses were generated with the resolution of 1 ~ 5 cm$^{-1}$. For carbon materials, the characteristic wave at the wave number of 1590 cm$^{-1}$ represents the sp$^2$-bonding graphite crystal layer (G band), while the wave peak at the wave number of 1350 $cm^{-1}$ symbolizes the $sp^3$-bonding defective structure (D band). The G band and the D band correspond to the conducting and the insulating structures, respectively. Using the lowest intensity count between these two peaks as the reference point, the relevant structure intensities ($I_G$, $I_D$) were obtained by integrating the curve along the wave number. The intensity ratio of $I_G/(I_D+I_G)$ signifies the degree of graphitization of CNTs. Therefore, the degree of graphitization is an indicator of electrical conductivity of CNTs. The higher the $I_G/(I_D+I_G)$ value is, the higher the degree of graphitization and thus the electrical conductivity of the CNTs are.

### 3. RESULTS AND DISCUSSION

With the lithographic and etching processes, the interconnect vias were patterned for CNT growth on the Ta bottom electrode, as shown in figure 2. At the conditions of $P_P$ = 1200 W, $P_G$ = 800 W, T = 600 °C, the smooth Ni film was transformed into discrete islands by hydrogen pretreatment. After methane of 30% flow ratio was introduced into the chamber for 15 minutes, well aligned CNTs in vias were synthesized, as presented in figure 3. The alignment is primarily induced by the electrical self-bias imposed on the substrate surface from the microwave plasma. Field-emission scanning electron microscopy (FESEM) was employed to inspect the morphology of Ni layer after hydrogen pretreatment and CNT films as well as measure the CNT diameters. The diameters of the CNTs range from about 32 to 40 nm, which basically conform the nanoparticle dimensions. The intensity ratio of $I_G/(I_D+I_G)$, was determined to be 0.445 from the corresponding Raman spectrum. The Ta top electrodes were deposited through sputtering method by applying the shadow mask on the CNT-grown substrates. Consequently, the CNT via diode devices were fabricated for the current (I)-voltage (V) electrical measurements.

Following similar procedures, the CNT diameters, graphitization indices, and the I-V characteristics of the vias were obtained for other process conditions on Ti and Ta metal electrodes, respectively. Table II indicates that the diameters of MWNTs grown on the bottom electrode of Ta decrease with increasing pretreatment power. It is caused from the smaller Ni nanoparticles produced by higher hydrogen pretreatment power. The corresponding Raman spectra are illustrated in figure 4(a). Both D band ($sp^3$-bonding defective structure) and G band ($sp^2$-bonding graphite crystal layer) are indicated at the characteristic wave numbers of 1350 $cm^{-1}$ and 1590 $cm^{-1}$, respectively. Figure 4(b) denotes that the effect of the pretreatment power on the CNT graphitization is negligible. It implies that the pretreatment power can only affect the nanoparticle dimensions and hence the CNT diameters. Figures 5 (a) and (b) present the current-voltage (I-V) characteristics of the CNT diodes of 0.35 μm 6x6 array vias on the bottom electrodes of Ta and Ti, respectively. Based on the linear attributes of the I-V curves, the resistances of the CNT diodes were calculated by the Ohm's law. These resistances result from the CNT resistances themselves and the contact resistances between CNTs and the metal electrodes. It is found that the CNT diode resistance decreases with the preatment power, as shown for 0.35 μm single via in figure 6. It is inferred that the decrease of the CNT diode resistance arises from the decrease of the CNT diameters.

Table III indicates that the diameters of MWNTs grown on the bottom electrode of Ta decrease with increasing substrate temperature while the effects of the growth power and methane flow ratio are insignificant. Figures 7 (a) and (b) reveal that the graphitization degree of MWNTs increases with growth plasma power and substrate temperature, respectively. Higher microwave power and thermal energy decompose the $CH_4$ gas efficiently and create a carbon-rich environment in the hydrocarbon-based plasma. This favors the formation of $sp^2$-like graphitic structures and hence leads to a better graphitization degree of CNTs. MWNT graphitization increases with methane flow ratio and begins to drop at the methane flow ratio of 30%, as shown in figure 7(c). It is inferred that the $CH_4$ flow ratio of less than 30% can not provide sufficient carbon feedstock for CNT growth. However, oversupply of methane (30%) produces a hydrogen-rich condition which is unfavorable to CNT formation. This is likely because the $sp^2$ carbons are attacked by hydrogen radicals and become $sp^3$ structures. The diameters of MWNTs grown on Ta bottom electrode are a little larger than those on Ti bottom electrode. This is attributed to the surface energy difference between Ta and Ti with respect to the catalytic film of nickel. At the conditions of $P_P$=1200 W, $P_G$=1200 W, T=600 °C, and methane flow ratio of 30%, the diameters have the largest discrepancy, as shown in Table IV. Figures 8 and 9 represent the variations of the CNT diode resistances of single via and array vias for 3.85x3.85 $\mu m^2$ and 2.45x2.45 $\mu m^2$ via regions, respectively. The square and circle symbols correspond to the CNT diodes with Ta and Ti bottom electrodes, respectively. The dashed and solid lines represent single via and array vias. Both figures have the same trends. The diode resistance of MWNTs in both single and array vias decreases with increasing MWNT graphitization degree. In the same via region, the MWNT diode resistances of the array vias are lower than those of the single vias. This can be ascribed to less tube–tube junctions and thus lower electrical resistance in the array vias [4]. The MWNT diode resistances on the bottom electrode of Ti are lower than those on the bottom electrode of Ta. It may be attributed to the smaller tube diameters on the bottom electrode of Ti and the work function difference between Ta and Ti films with respect to the work function of CNTs. The 0.35 µm 4x4 array vias can be considered as 16 0.35 µm single vias connected in parallel, i.e. $R_{0.35\ \mu m\ 4x4\ array\ vias}$ (actual) = $R_{0.35\ \mu m\ single\ via}$ /16 (Caclulated). The great consistency between the measured CNT diode resistances of 4x4 array vias and the calculated ones from the single via is illustrated in figure 10. It confirms the concept of CNT array vias.

## REFEERNCES

Table I Process parameters of CNTs grown by MPCVD

| $P_P$ (watts) | $P_G$ (watts) | Temp. ($^oC$) | $[CH_4/(H_2+CH_4)]$ (%) |
|---|---|---|---|
| **800** | 800 | 600 | 30 |
| **1000** | 800 | 600 | 30 |
| **1200** | 800 | 600 | 30 |
| 1200 | **800** | 600 | 30 |
| 1200 | **1000** | 600 | 30 |
| 1200 | **1200** | 600 | 30 |
| 1200 | 800 | **550** | 30 |
| 1200 | 800 | **600** | 30 |
| 1200 | 800 | **650** | 30 |
| 1200 | 800 | 600 | **20** |
| 1200 | 800 | 600 | **30** |
| 1200 | 800 | 600 | **40** |

Table II Variation of MWNT diameters (D) with pretreatment plasma power on the bottom electrode of Ta ($P_G$=800 W, T=600 ℃, $CH_4$ flow ratio=30%)

| $P_P$ (W) | 800 | 1000 | 1200 |
|---|---|---|---|
| $D_{Range}$ | 36-47 | 33-50 | 32-40 |
| $D_{Average}$ | 42.7 | 39.7 | 36 |

Table III Variation of MWNT diameters (D) with substrate temperature on the bottom electrode of Ta ($P_P$=1200 W, $P_G$=800 W, $CH_4$ flow ratio=30%)

| T ($^oC$) | 550 | 600 | 650 |
|---|---|---|---|
| $D_{Range}$ | 45-59 | 32-40 | 19-28 |
| $D_{Average}$ | 52.5 | 36 | 23.7 |

Table IV Difference of MWNT diameters (D) between bottom electrodes Ti and Ta ($P_P$=1200 W, $P_G$=1200 W, T=600 $^oC$, $CH_4$ flow ratio=30%)

| Bottom electrode | Ti | Ta |
|---|---|---|
| $D_{Range}$ | 24-32 | 32-40 |
| $D_{Average}$ | 28.1 | 36 |

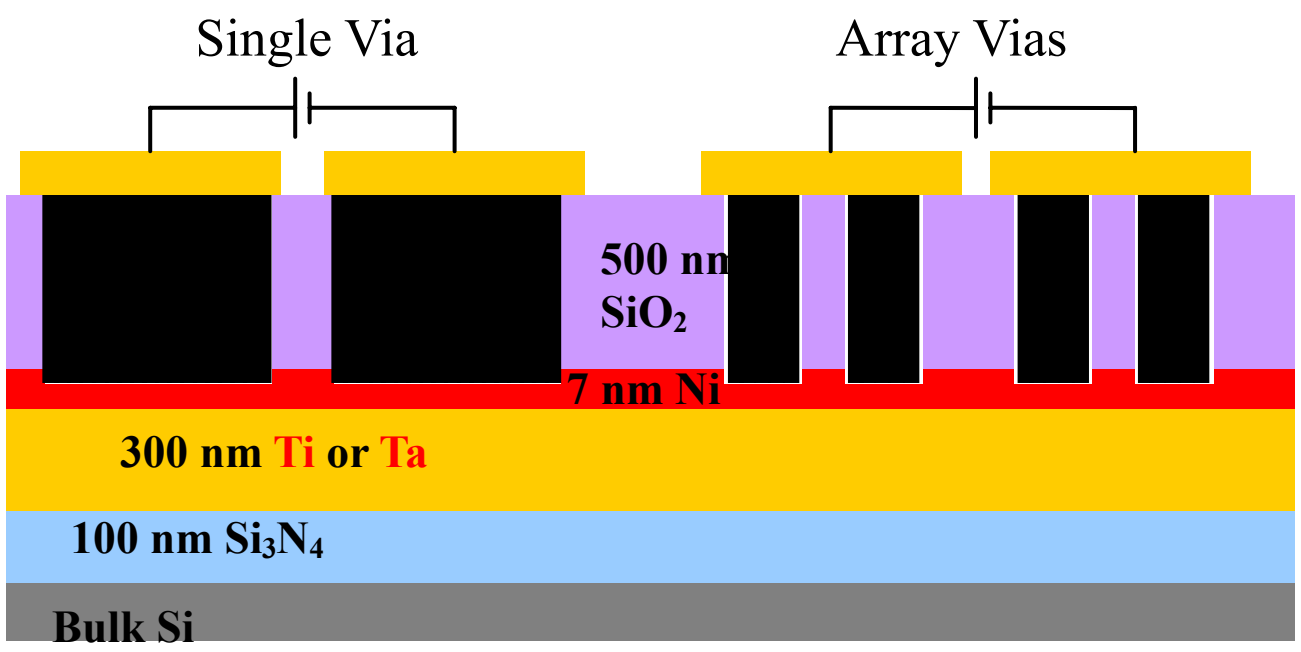

Figure 1 Cross section of the CNT diode structure

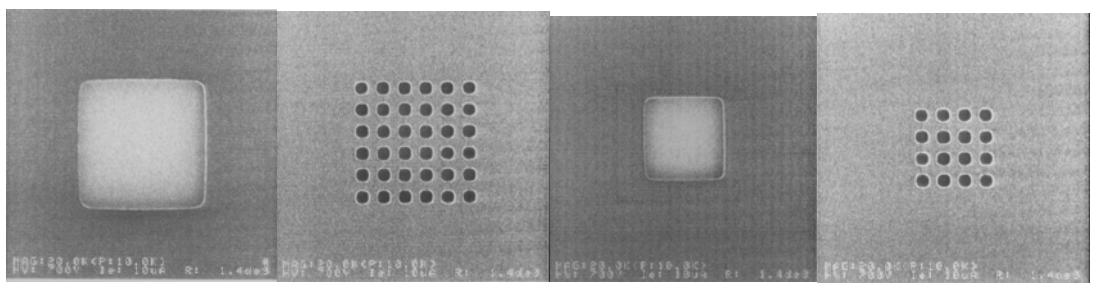

(a) 3.85x3.85 μm$^2$ (b) 2.45x2.45 μm$^2$ via regions

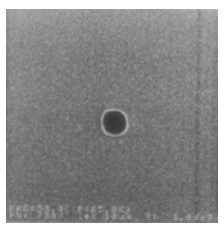

(c) 0.35 μm single via

Figure 2 SEM images of single-via and array-via patterns

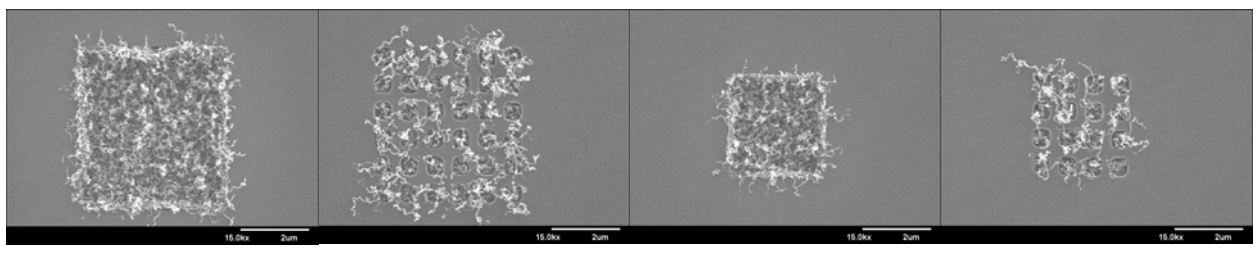

(a) 3.85 μm (b) 6x6 array (c) 2.45 μm (d) 4x4 array

Figure 3 SEM images of MWNTs grown in the vias by MPCVD

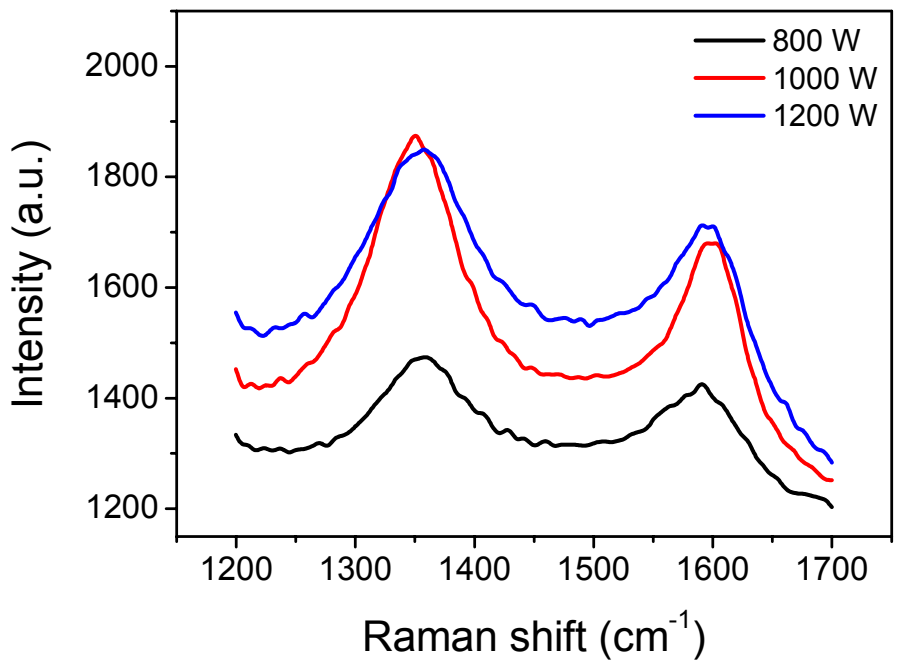

(a) Raman spectra

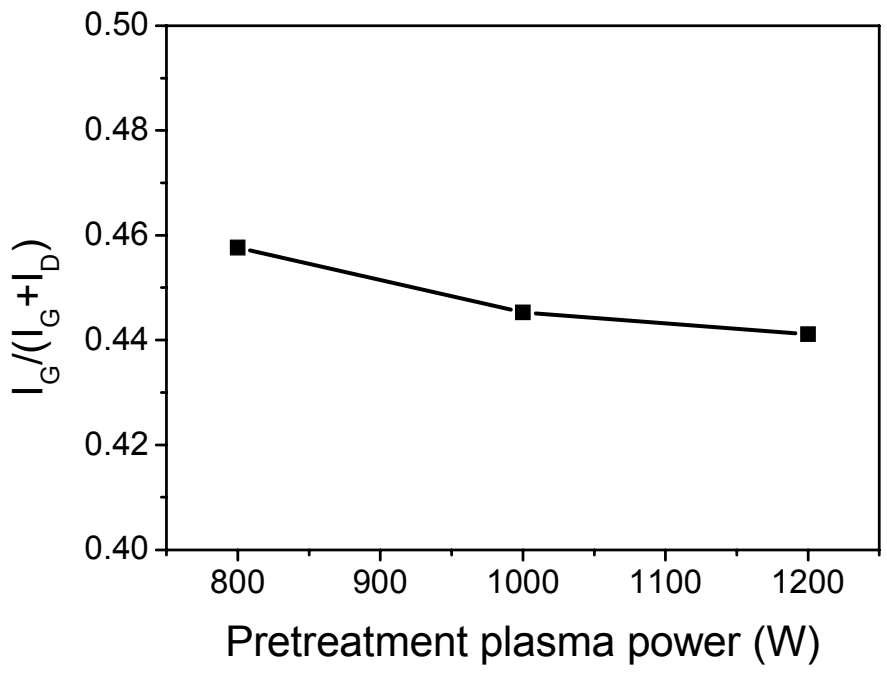

(b) Intensity ratio of CNT graphitization

Figure 4 Variation of MWNT graphitization with pretreatment power ($P_G$=800 W, 600 °C, 30%)

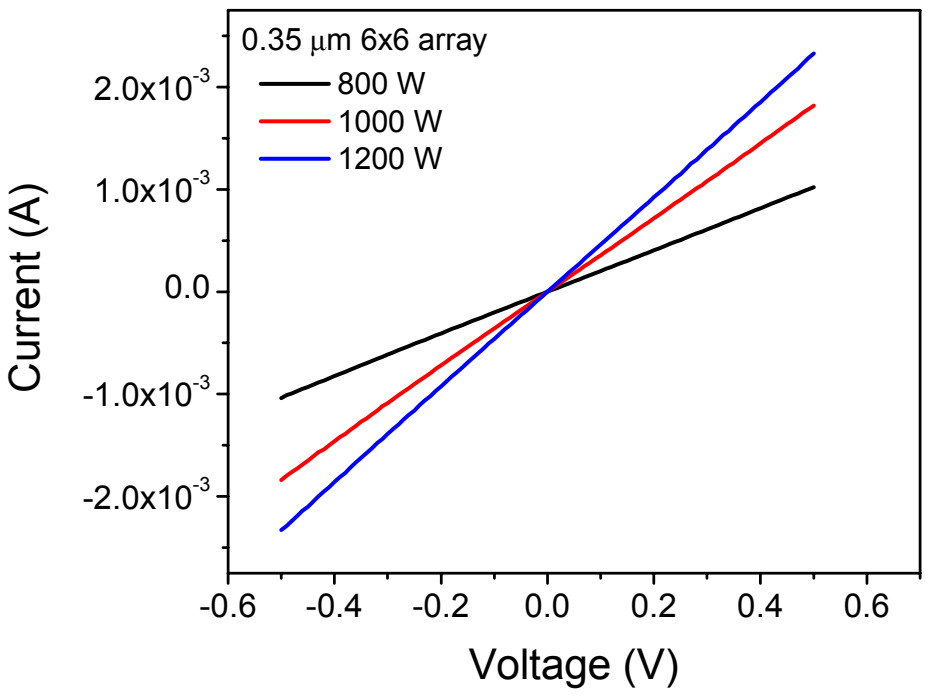

(a) Ta bottom electrode

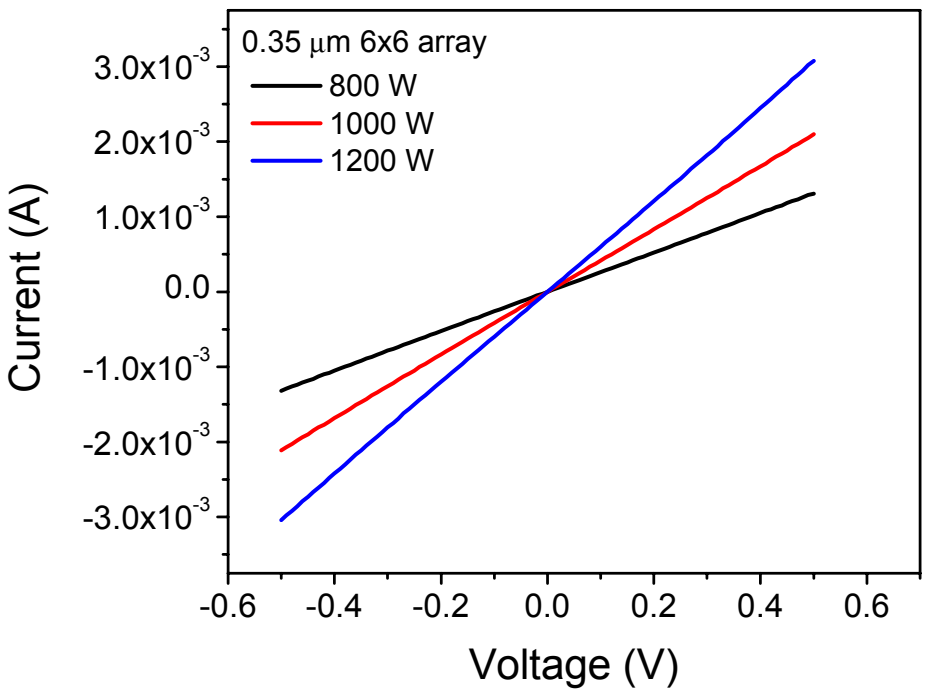

(b) Ti bottom electrode

Figure 5 I-V characteristics of CNT diodes of 0.35 μm 6x6 array vias ($P_G$=800 W, 600 °C, 30%)

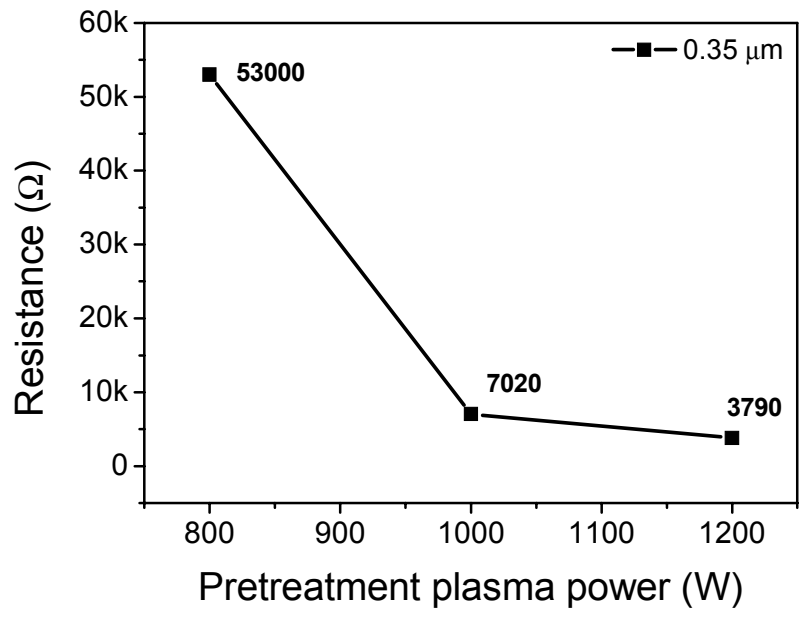

(a) Ta bottom electrode

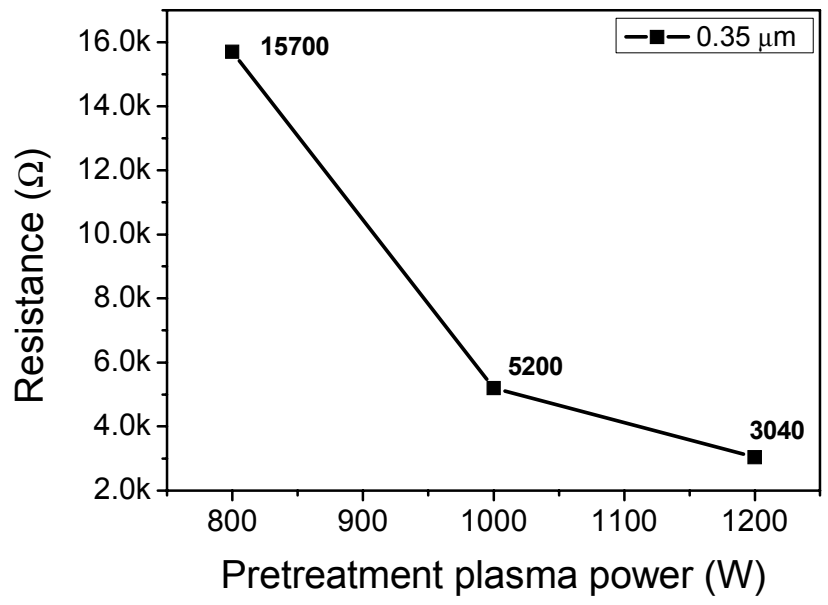

(b) Ti bottom electrode

Figure 6 Variation of the CNT diode resistance with the preatment power, for 0.35 μm single via ($P_G$=800 W, 600 °C, 30%)

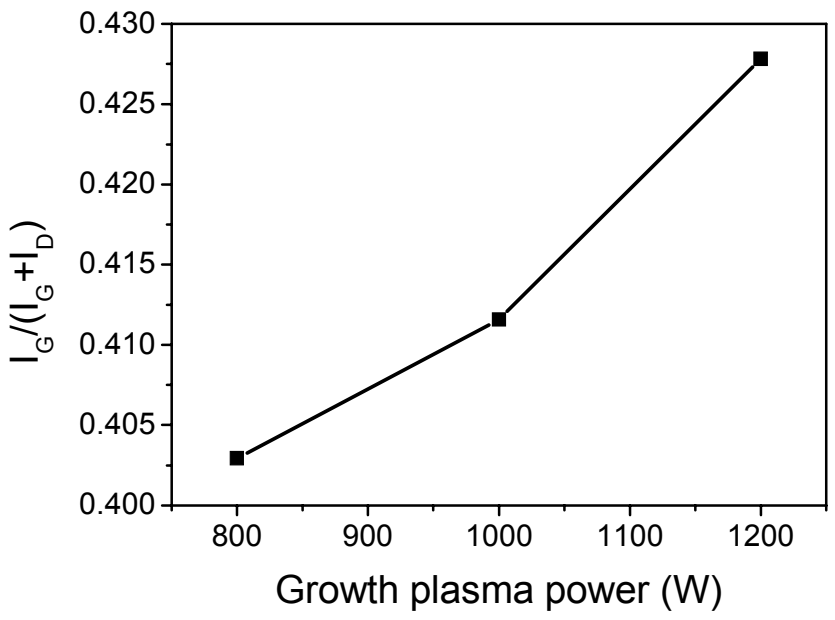

(a) Growth plasma power

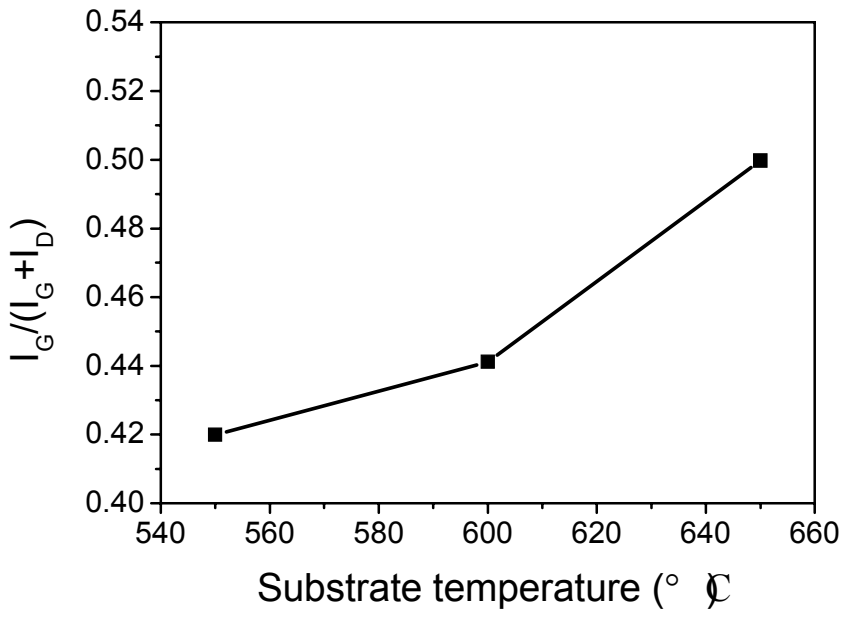

(b) Substrate temperature

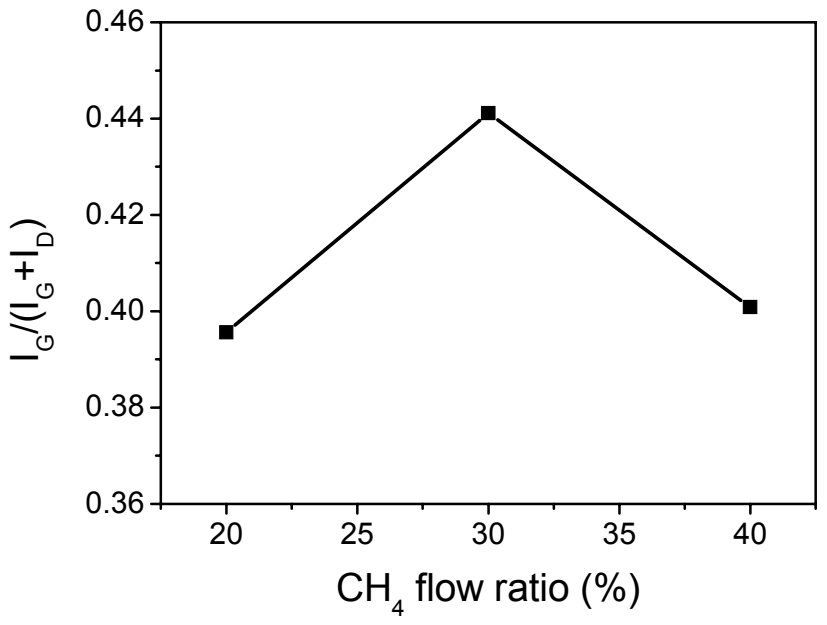

(c) Methane flow ratio

Figure 7 Dependence of CNT graphitization on the process paramenters

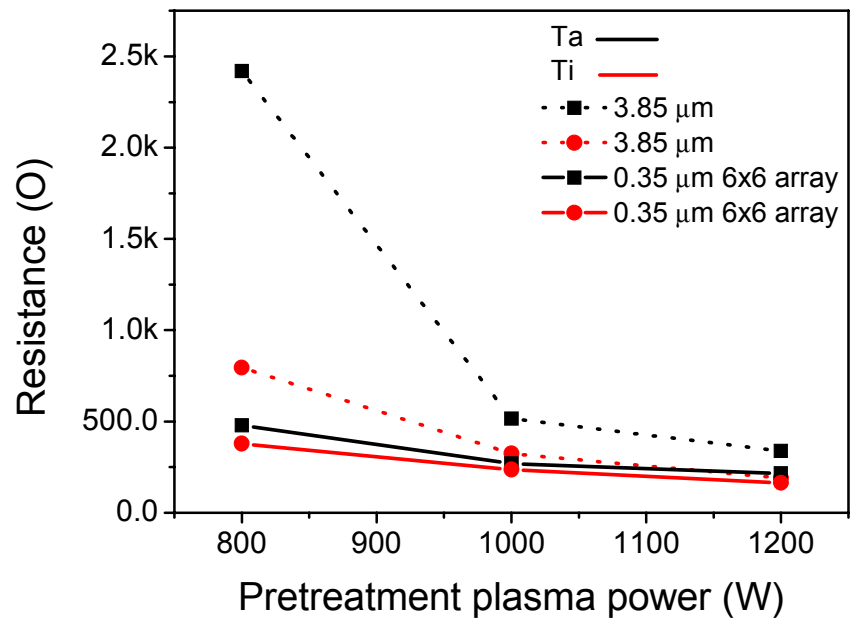

(a) Pretreatment power

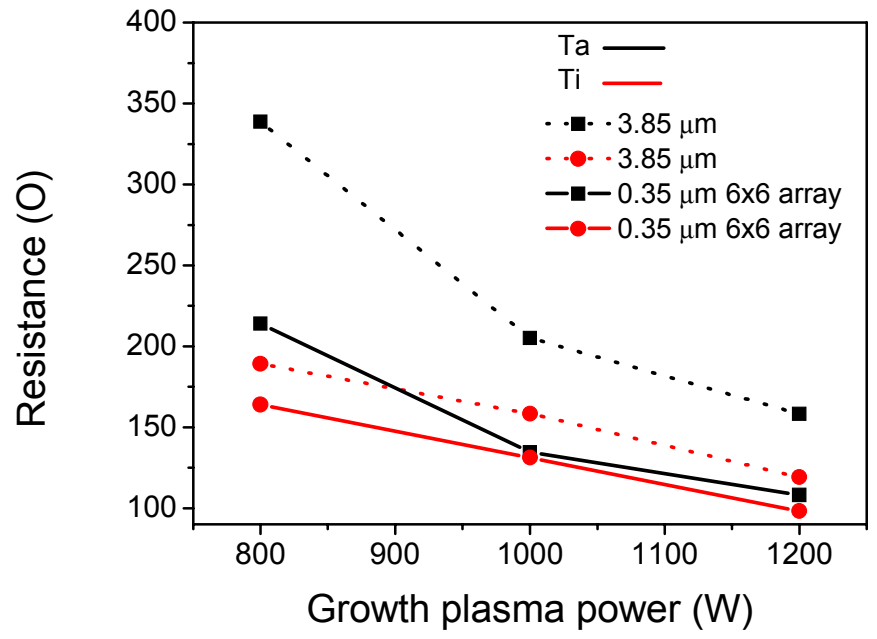

(b) Growth power

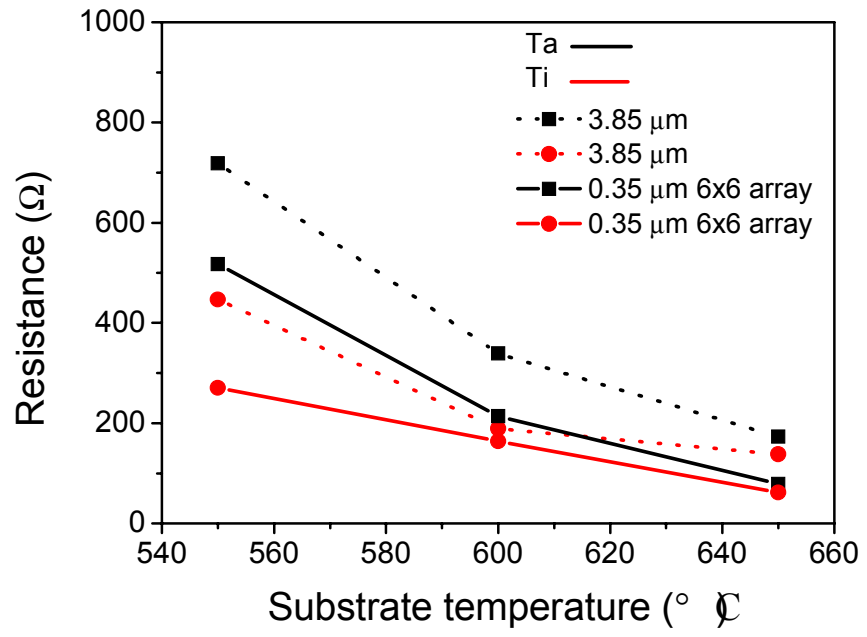

(c) Substrate temperature

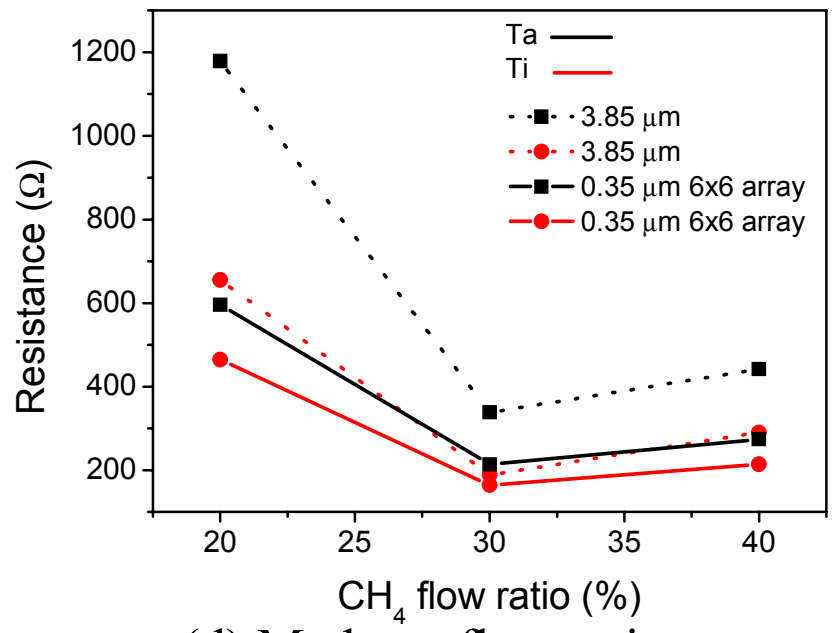

(d) Methane flow ratio

Figure 8 Diode resistance of MWNTs in the same via region (3.85 μm) with bottom electrodes of Ta and Ti

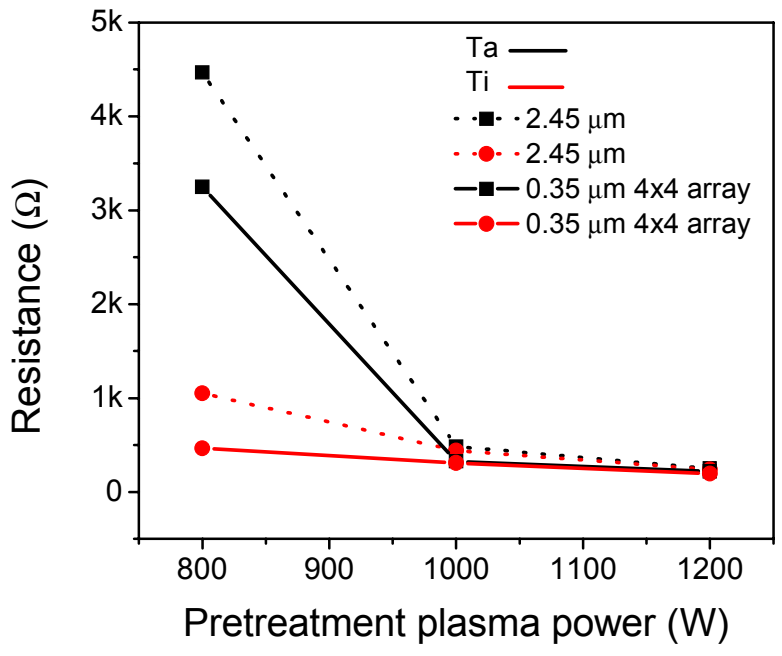

(a) Pretreatment power

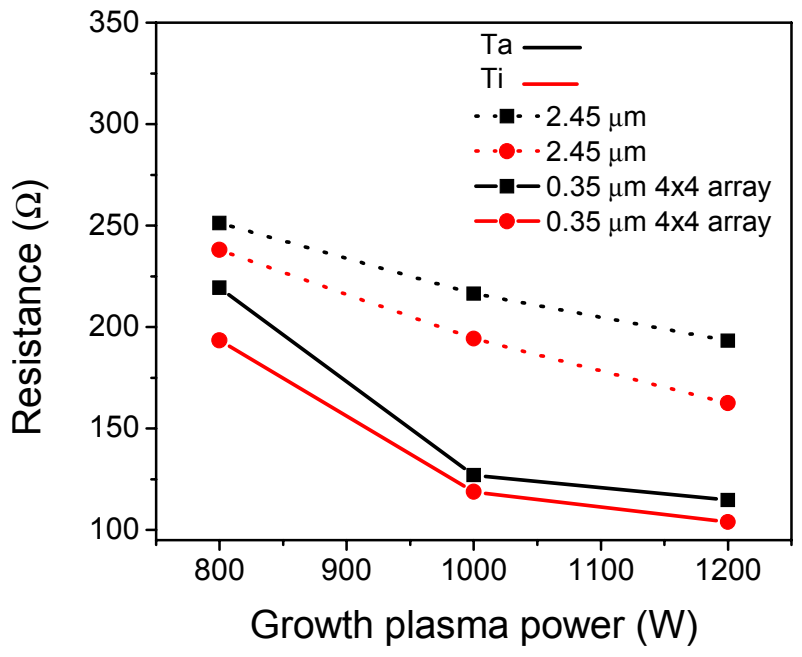

(b) Growth power

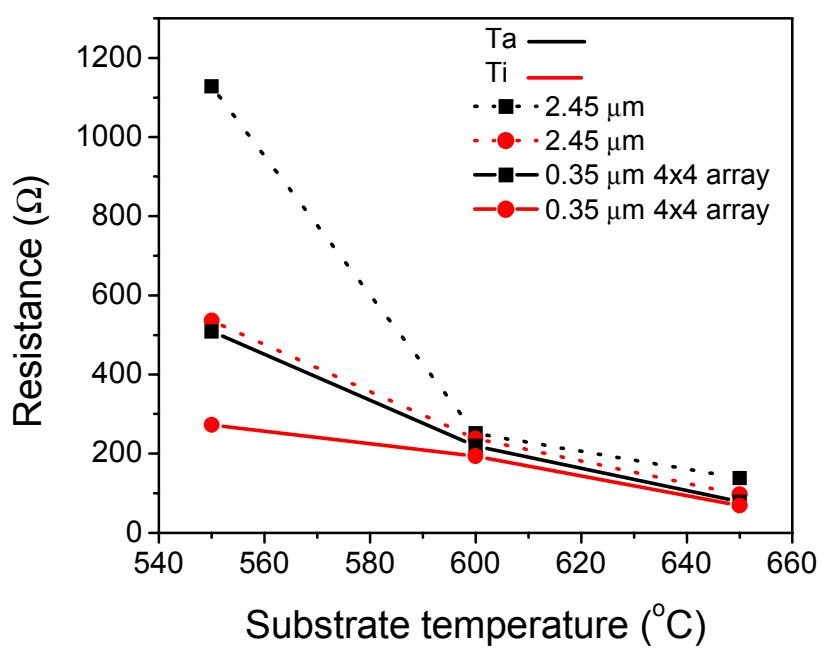

(c) Substrate temperature

Figure 9 Diode resistance of MWNTs in the same via region (2.45 µm) with bottom electrodes of Ta and Ti

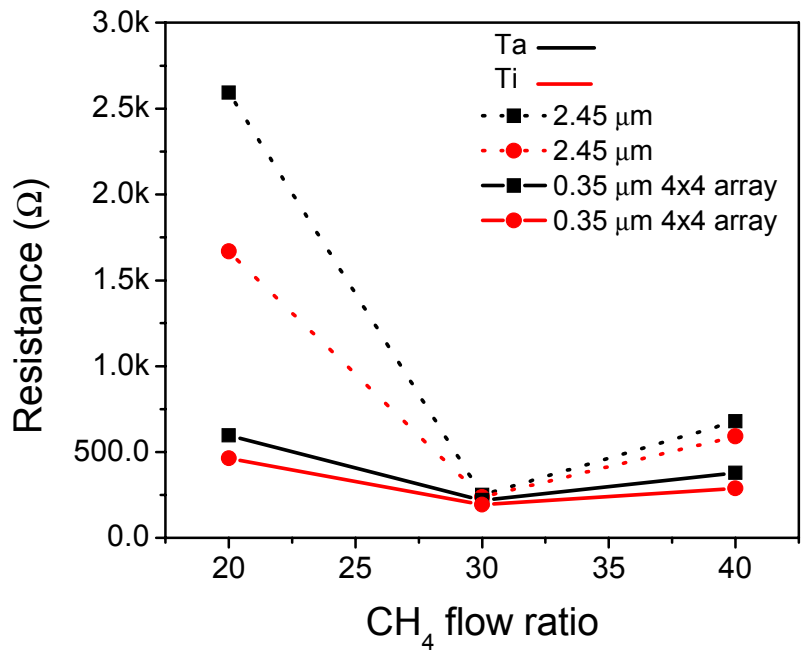

(d) Methane flow ratio

Figure 9 Diode resistance of MWNTs in the same via region (2.45 µm) with bottom electrodes of Ta and Ti (Concluded)

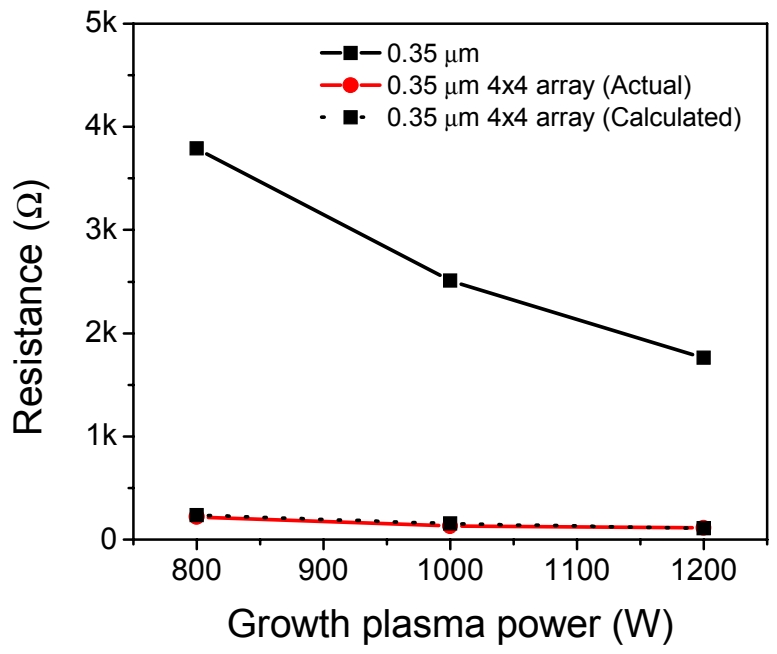

Figure 10 The CNT diode resistances of 4x4 array vias and the calculated ones from 0.35 µm single via